\documentclass[runningheads,a4paper,table,11pt]{llncs}
\usepackage[utf8]{inputenc}
\usepackage{amsmath,amssymb,amsfonts}
\usepackage{algpseudocode,algorithm,algorithmicx}
\usepackage{graphicx}
\usepackage{textcomp}
\usepackage{parskip}
\usepackage{enumitem}
\usepackage{subcaption}
\usepackage{lipsum}
\usepackage{url}
\usepackage{lineno}

\input{uppaal.sty}

\usepackage{placeins}

\usepackage{multirow}
\usepackage{tabularx}
\newcolumntype{Y}{>{\centering\arraybackslash}X}

\usepackage{tikz}
\usepackage{environ}
\usetikzlibrary{automata, positioning, arrows, fit, shapes, backgrounds, calc}

\newcommand{\convexpath}[2]{
[   
    create hullnodes/.code={
        \global\edef\namelist{#1}
        \foreach [count=\counter] \nodename in \namelist {
            \global\edef\numberofnodes{\counter}
            \node at (\nodename) [draw=none,name=hullnode\counter] {};
        }
        \node at (hullnode\numberofnodes) [name=hullnode0,draw=none] {};
        \pgfmathtruncatemacro\lastnumber{\numberofnodes+1}
        \node at (hullnode1) [name=hullnode\lastnumber,draw=none] {};
    },
    create hullnodes
]
($(hullnode1)!#2!-90:(hullnode0)$)
\foreach [
    evaluate=\currentnode as \previousnode using \currentnode-1,
    evaluate=\currentnode as \nextnode using \currentnode+1
    ] \currentnode in {1,...,\numberofnodes} {
-- ($(hullnode\currentnode)!#2!-90:(hullnode\previousnode)$)
  let \p1 = ($(hullnode\currentnode)!#2!-90:(hullnode\previousnode) - (hullnode\currentnode)$),
    \n1 = {atan2(\y1,\x1)},
    \p2 = ($(hullnode\currentnode)!#2!90:(hullnode\nextnode) - (hullnode\currentnode)$),
    \n2 = {atan2(\y2,\x2)},
    \n{delta} = {-Mod(\n1-\n2,360)}
  in 
    {arc [start angle=\n1, delta angle=\n{delta}, radius=#2]}
}
-- cycle
}

\usepackage{varioref}
\usepackage{hyperref}

\makeatletter
\newsavebox{\measure@tikzpicture}
\NewEnviron{scaletikzpicturetowidth}[1]{%
  \def\tikz@width{#1}%
  \def\tikzscale{1}\begin{lrbox}{\measure@tikzpicture}%
  \BODY
  \end{lrbox}%
  \pgfmathparse{#1/\wd\measure@tikzpicture}%
  \edef\tikzscale{\pgfmathresult}%
  \BODY
}
\makeatother

{\dimen0=#1\relax
 \dimen1=\dimexpr 0.5\dimen0 - 0.5\textwidth\relax
 \hspace*{-\dimen1}\minipage{\dimen0}}%
{\endminipage\hspace{-\dimen1}}

\begin{document}
\title{Near Optimal Task Graph Scheduling with Priced Timed Automata
  and Priced Timed Markov Decision Processes}






\author{Anne Ejsing \and Martin Jensen \and Marco Muñiz \and Jacob
  Nørhave\and Lars~Rechter} \institute{Aalborg University, Aalborg,
  Denmark \email{\{aejsin,martje,jnarha,lrecht\}16@student.aau.dk,
    muniz@cs.aau.dk}}


\subtitle{\bf\large Technical Report}

\maketitle

\begin{abstract}
  Task graph scheduling is a relevant problem in computer science with
  application to diverse real world domains. Task graph scheduling
  suffers from a combinatorial explosion and thus finding optimal
  schedulers is a difficult task.
  In this paper we present a methodology for computing near-optimal
  preemptive and non-preemptive schedulers for task graphs. The task
  graph scheduling problem is reduced to location reachability via the
  fastest path in Priced Timed Automata (PTA) and Priced Timed Markov
  Decision Processes (PTMDP). Additionally, we explore the effect of
  using chains to reduce the computation time for finding schedules.
  We have implemented our models in \uppaalcora and \stratego.  We
  conduct an exhaustive experimental evaluation where we compare our
  resulting schedules with the best-known schedules of a state of the
  art tool.  A significant number of our resulting schedules are shown
  to be shorter than or equal to the best-known schedules. 
\end{abstract}


\section{Introduction and Motivation}
The task graph scheduling problem is a well known and widely discussed
problem in mathematics and computer science. Creating optimal and
near-optimal schedules are relevant in many real-world applications,
such as scheduling of computations in spreadsheets for parallel
execution \cite{Guldstrand2018_unpub}.
Given a task graph, computing an optimal schedule is NP-complete
\cite{Kwok1999}. Applications such as spreadsheets, induce large task
graphs and computation of optimal schedulers in such domains is
intractable. Therefore, we investigate how to obtain a near-optimal
solution. Furthermore, we add preemption to find shorter schedules. As
the task graph scheduling problem is reducible to the reachability
problem in timed automata~\cite{Abdeddaim2003TaskGraphScheduling}, we
use extensions of timed automata to find near-optimal solutions, being
Priced Timed Automata and Priced Timed Markov Decision Processes.

Our main contribution is a methodology for modelling task graphs, with
preemptive and non-preemptive schedulers using priced timed automata
and priced timed markov decision processes. The intuition of our
preemptive models is that we preempt all executing tasks when any of
the executing tasks have finished. In this work we assume that
preemption has no cost, however an arbitrary cost can be easily
implemented. We thoroughly evaluate our approach on the standard task
graph set of Kasahara and Narita~\cite{kasahara_schedule}, which
provides the best-known schedules. Furthermore, we compare schedules
that allow preemption, to schedules that do not. This is to conclude
when it is beneficial to use preemption. Lastly, we evaluate our
implementations of our models based on priced timed automata and
priced timed markov decision processes to investigate when either is
beneficial.

\subsubsection*{Related Work}
\label{sec:RelatedWork}
In this article, we use extensions of timed automata to find task
graph schedules. Timed automata originates from the work of Alur and
Dill~\cite{AlurDill1994}.
%
%
In~\cite{Abdeddaim2003TaskGraphScheduling} Abdeddaïm, Kerbaa, and
Maler use timed automata to compute schedulers for task graphs. They
do so non-preemptively and they compose these automata using a special
parallel mutual exclusion composition operator. They compute a
schedule via location reachability in the resulting automata.
The work of Kasahara and
Narita~\cite{kasahara_schedule,Kasahara_article} comprises a standard
task graph set and the best schedules they found for each of them. To
obtain these non-preemptive schedules they use a branch and bound
algorithm. In our work, we use the standard task graph set and the
best-known schedules of Kasahara and Narita~\cite{kasahara_schedule},
to evaluate the quality of the schedules we obtain with our models.
In~\cite{Guldstrand2018_unpub} the authors show an attempt of
analyzing spreadsheets for parallel execution via model checking. The
authors focus on optimal schedulers and do not use preemption. They
explore only small and few task graphs.
Contrary to the work listed, we model task graphs both preemptively
and non-preemptively. Besides we use priced timed automata and priced
timed markov decision processes, implemented in \uppaalcora and
\stratego, respectively. This is to explore more options for achieving
shorter schedules. Lastly, we achieve near-optimal results on larger
task graphs, rather than optimal results on small task graphs as
in\cite{Guldstrand2018_unpub}.

\subsubsection*{Outline}
This paper is structured as follows;
Section~\ref{sec:Task_Graph_Scheduling} introduces the theory behind
task graph scheduling and preemption. Section~\ref{sec:chain_theory}
introduces the notion of chain decomposition of a task
graph. Following this, Section~\ref{sec:Timed_Automata} presents the
relevant theory of priced timed automata and priced timed markov
decision processes. The implementation and methodology for modelling
our models in \uppaalcora and \stratego is documented in
Section~\ref{sec:UppaalModels}. In Section~\ref{sec:tests_results} we
compare the schedules obtained by our approach with the best-known
schedules.



\section{Task Graph Scheduling}
\label{sec:Task_Graph_Scheduling}

In this section, we define \emph{task graphs} and \emph{schedules} in
the style
of~\cite{Abdeddaim2003TaskGraphScheduling,Abdeddaim2006SchedulingWithTA},
we also introduce the notion of preemption in this context.
Task graph scheduling is the activity of assigning tasks to
machines. The result of task graph scheduling is a \emph{schedule}
which is a function that maps tasks to execution start times and
durations.


\begin{definition}[Task Graph]
  A task graph is a triple:
  $(\mathcal{P}, \sqsubset, D)$ where
  $\mathcal{P} = \{P_1,...,P_i\}$ is the set of all tasks, $\sqsubset$
  is a strict partial-order relation on the set of tasks, and $D$ is a
  function, $D: P \rightarrow \mathbb{N}$ that assigns duration to
  each of the tasks of P.
  \label{def:TaskGraph}
\end{definition}
 
Let M be a set of $j$ identical machines, $M = \{m_1,...,m_j\}$. The
problem consists of assigning tasks to machines in periods of time
such that:
\begin{itemize}[noitemsep,topsep=0pt]
\item A task can run iff all of its predecessors are completed.
\item Each machine can run at most one task at a time.
\end{itemize}

Such an assignment is known as a schedule. Schedules can be created
either preemptively or non-preemptively.


\begin{definition}[Possible and Optimal Schedules]
  \label{def:Schedules} For a task graph ($\mathcal{P}$, $\sqsubset$,
D) and $j$ machines, a possible schedule is a function $start:
\mathcal{P} \rightarrow S$ where $S \subseteq \mathbb{R}_{\geq0}\times
\mathbb{R}_{>0}$ is a set of pairs $(s,d)$, denoting respectively
start times and intermediate durations of the task. The function must
satisfy:
  \begin{itemize}
  \item No task can run before all of its predecessors have completed,
formally: For every $P, P' \in \mathcal{P}$ where $P' \sqsubset P$
then $inf(\{s\ |\ (s,d) \in start(P)\}) \geq sup(\{s'+d'\ |\
(s',d')\in start(P')\})$
  \item At any point in time at most j machines can run tasks,
formally: For any $t \in \mathbb{R^+}$ then $|\{P\ |\ P \in
\mathcal{P}\ \text{and}\ (s,d) \in start(P)\ and\ {t} \cap [s,s+d]
\neq \emptyset\}| \leq j$
\item Running a single task in parallel is not allowed, formally: For
  every two pairs $(s,d), (s',d') \in start(P)$ where
  $P \in \mathcal{P}$ and $s\neq s'$ or $d\neq d'$ then either
  $s+d \leq s'$ or $s'+d' \leq s$.
  \item The sum of the intermediate durations of a task is equal to
the full duration of the task, formally: For every $P \in \mathcal{P}$
then $\sum_{(s,d) \in start(P)} d = D(P)$
\end{itemize} The length of a schedule is the time when the last task
finishes execution, formally;
$sup(\{s+d\ |\ (s, d) \in start(P)\ \text{for every}\ P\ in\
\mathcal{P}\})$. An optimal schedule has the minimal length. 
\end{definition}

Note that the set of non-preemptive schedules is a subset of that of
preemptive schedules where the schedule contains only one pair $(s,d)$
for each task.

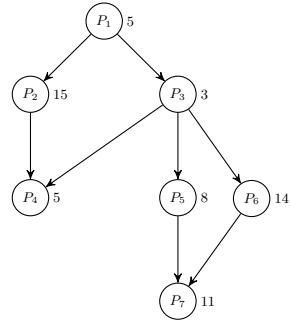
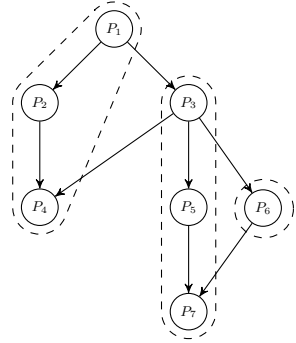
\begin{figure}[t]
  \centering
  \begin{subfigure}[t]{0.38\textwidth}
    \centering
    \begin{scaletikzpicturetowidth}{0.8\textwidth}
    \begin{tikzpicture}[scale=\tikzscale,->,>=stealth',node distance=1.5cm,transform shape]
        \node[state] (q1) {$P_1$};
        \node[state, below left=of q1] (q2) {$P_2$};
        \node[state, below right=of q1] (q3) {$P_3$};
        \node[state, below=of q2] (q4) {$P_4$};
        \node[state, below=of q3] (q5) {$P_5$};
        \node[state, below right=of q3, yshift=-7mm] (q6) {$P_6$};
        \node[state, below=of q5] (q7){$P_7$};
        
        \node[right=of q1, xshift=-15.2mm] {\footnotesize $5$};
        \node[right=of q2, xshift=-15.2mm] {\footnotesize $15$};
        \node[right=of q3, xshift=-15.2mm] {\footnotesize $3$};
        \node[right=of q4, xshift=-15.2mm] {\footnotesize $5$};
        \node[right=of q5, xshift=-15.2mm] {\footnotesize $8$};
        \node[right=of q6, xshift=-15.2mm] {\footnotesize $14$};
        \node[right=of q7, xshift=-15.2mm] {\footnotesize $11$};
        
        \draw   (q1) edge (q2)
                (q2) edge (q4)
                (q1) edge (q3)
                (q3) edge (q4)
                (q3) edge (q5)
                (q3) edge (q6)
                (q5) edge (q7)
                (q6) edge (q7);
    \end{tikzpicture}
    \end{scaletikzpicturetowidth}
    \caption{Example of a task graph}
    \label{fig:Task-Graph}
\end{subfigure}~
  \\
  \begin{subfigure}[t]{0.38\textwidth}
    \centering
    \begin{scaletikzpicturetowidth}{0.8\textwidth}
    \begin{tikzpicture}[scale=\tikzscale,->,>=stealth',node distance=1.5cm,transform shape]
        \node[state] (q1) {$P_1$};
        \node[state, below left=of q1] (q2) {$P_2$};
        \node[state, below right=of q1] (q3) {$P_3$};
        \node[state, below=of q2] (q4) {$P_4$};
        \node[state, below=of q3] (q5) {$P_5$};
        \node[state, below right=of q3, yshift=-5.4mm, xshift=-1.5mm, dashed, minimum size=1.3cm] (q6_1) {};
        \node[state, below right=of q3, yshift=-7mm] (q6) {$P_6$};
        \node[state, below=of q5] (q7){$P_7$};
        
        \draw   (q1) edge (q2)
                (q2) edge (q4)
                (q1) edge (q3)
                (q3) edge (q4)
                (q3) edge (q5)
                (q3) edge (q6)
                (q5) edge (q7)
                (q6) edge (q7);
                
        \draw[dashed] \convexpath{q1,q4,q2}{0.6 cm};
        \draw[dashed] \convexpath{q3,q7}{0.6 cm};
    \end{tikzpicture}
    \end{scaletikzpicturetowidth}
    \caption{Chain decomposition}
    \label{fig:Task-Graph_chains}
\end{subfigure}  
  \caption{Example of task graph and chain decomposition}
\end{figure}

\begin{figure}[t]  \centering
  \begin{subfigure}[b]{0.45\textwidth} 
    \includegraphics[width=\textwidth]{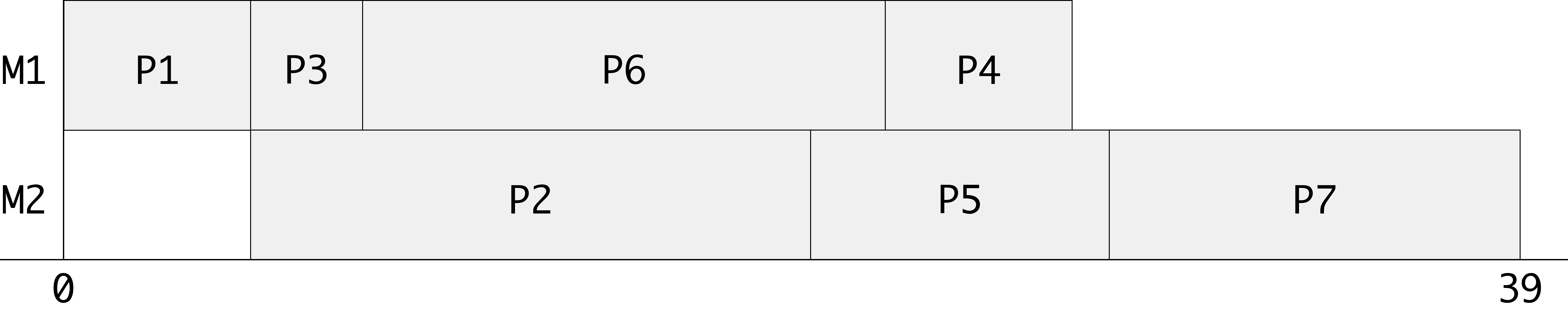}
    \caption{Intuitive non-preemptive schedule}
    \label{fig:intuitive_schedule}    
  \end{subfigure} 
  \begin{subfigure}[b]{0.45\textwidth}
    \includegraphics[width=\textwidth]{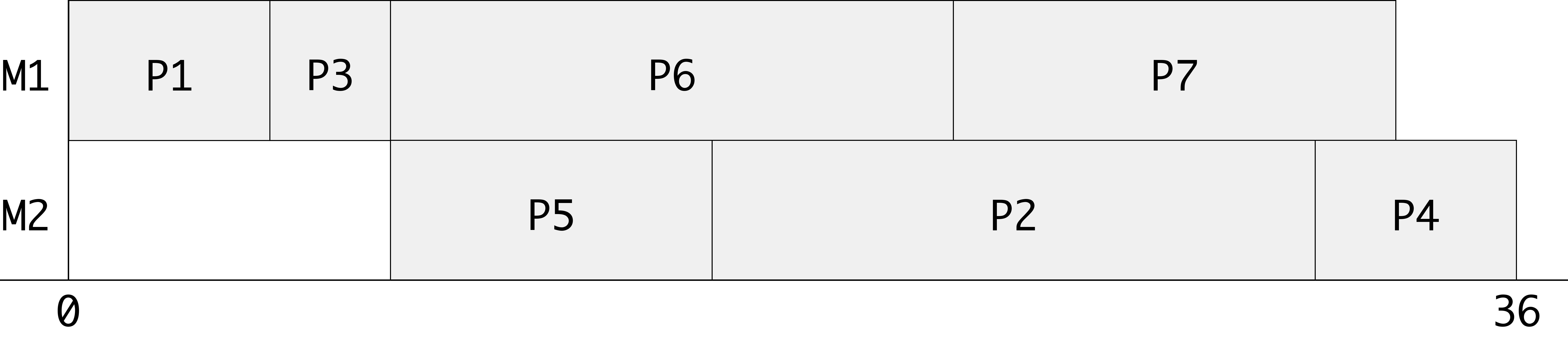}
    \caption{Optimal non-preemptive schedule}
    \label{fig:optimal_non_preemption}
  \end{subfigure}
  
  \begin{subfigure}[b]{0.45\textwidth}
    \includegraphics[width=\textwidth]{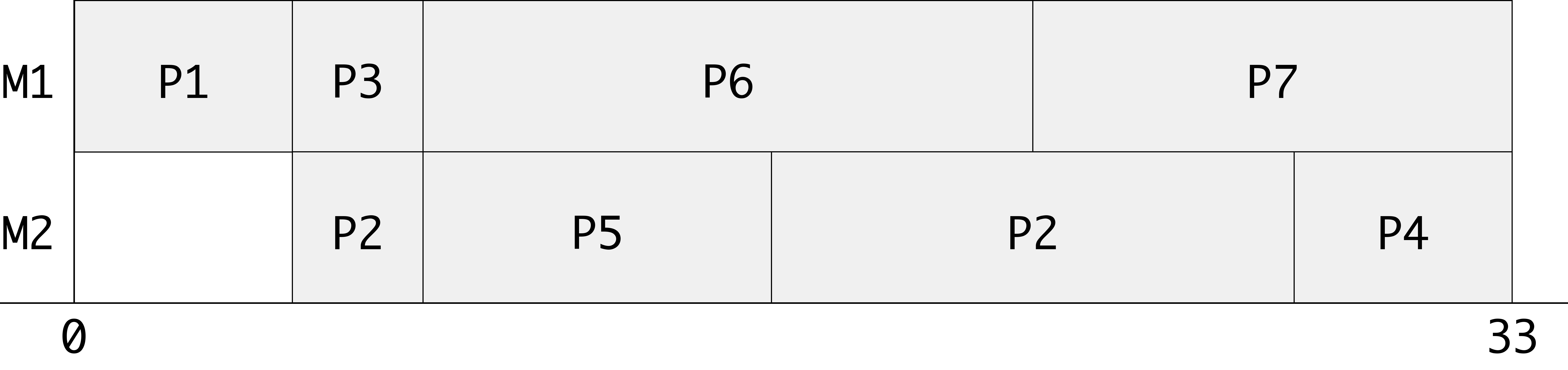}
    \caption{Optimal preemptive schedule}
    \label{fig:optimal_preemptive}
  \end{subfigure}
  \caption{Examples of schedules obtained using preemption and
    non-preemption.}
\end{figure}

The scheduler’s goal is to obtain a schedule of minimal length. If the
number of machines is greater than or equal to the number of tasks,
this is trivial; the scheduler assigns a machine to each task. With
fewer machines than tasks, the scheduler must decide which tasks to
run first. To demonstrate the complexity, consider Figure
\ref{fig:Task-Graph}. The intuitive non-preemptive way to schedule is
to assign tasks to machines as they become available (see Figure
\ref{fig:intuitive_schedule}). We achieve a shorter schedule when one
machine idles while the other computes task 3 (see Figure
\ref{fig:optimal_non_preemption}). Despite the optimal non-preemptive
scheduler reducing the length of the schedule, it still contains idle
time. To reduce this, we use preemption which results in a shorter
schedule than Figure \ref{fig:optimal_non_preemption}, which is seen
in Figure \ref{fig:optimal_preemptive}. Now, the only idle time
present is to fulfill the dependencies. The dilemma of the
non-preemptive scheduler (whether optimality is reached by computing a
task or idling) is now irrelevant. This is because, for the preemptive
scheduler, it is \emph{always} optimal to compute a task if
possible. This is based on our assumption that context-switching is
free. The task graph scheduling problem is a vast combinatorial
problem and is NP-complete \cite{Kwok1999}. We, therefore, explore how
to find \emph{near optimal schedules} rather than optimal ones.

\subsection{Scheduling with Chains}
\label{sec:chain_theory} To find all possible schedules, the scheduler
must consider all combinations of executing all tasks while still
fulfilling each task's dependencies. Thus, it must iterate through
every task of the task graph each time a new task is to be
scheduled. To reduce the number of tasks that the scheduler has to
iterate through, we adopt the use of
chains as in~\cite{Abdeddaim2006SchedulingWithTA}. A chain is a directly
connected route in a graph representation of a partially ordered
set. We formally define chains in Definition \ref{def:Chain}. We
define the decomposition of a partial order into chains in Definition
\ref{def:Chain_Decomposition}.

\begin{definition}[Chain]
  \label{def:Chain} A chain, $C$, in a strict partially ordered set,
$(\mathcal{P},\sqsubset)$, is a subset $C \subseteq \mathcal{P}$ such
that for any two tasks $P, P' \in C$ where $P \neq P'$ then $P
\sqsubset P'$ or $P' \sqsubset P$.
\end{definition}

\begin{definition}[Chain Decomposition]
  \label{def:Chain_Decomposition} A chain decomposition is a partition
of the elements of a partial order into chains.
\end{definition}

The decomposition of the task graph of Figure \ref{fig:Task-Graph} to
chains is seen in Figure \ref{fig:Task-Graph_chains}. The task graph
thus comprises three chains being $chain_1 = \{P_1,P_2,P_4\}$,
$chain_2 = \{P_3,P_5,P_7\}$, and $chain_3 = \{P_6\}$. Note that
$\sqsubset$ totally orders the elements in each of the chains, as the
definition of chains adds comparability to these subsets. As
$\sqsubset$ from Definition \ref{def:TaskGraph} describes a
dependency, then there is exactly one task in each chain that does not
depend on another task in the chain; the least element of the
chain. Once the task that is the least element is computed, we remove
it from its chain such that the chain has a new least element. This
way, the scheduler only needs to iterate through the least element of
each chain, rather than iterating through every task when finding the
tasks that can be scheduled. We state that we can do this without
limiting the scheduler from creating any possible schedule.

\begin{lemma}[Chain Optimality]
  \label{lemma:Chain_optimality} Given a task graph, any possible
schedule, and any chain decomposition, then the schedule can be
computed by considering only the least elements of the chains at any
point in time.
\end{lemma}

\subsubsection*{Computation of Chains}
\label{sec:chain_creation}
Note that due to Lemma~\ref{lemma:Chain_optimality} it is not required
to compute an optimal chain decomposition to preserve the best
possible schedule.
We have developed an algorithm which computes a chain cover of a given
task graph. 
The intuition of the algorithm is to use the number of predecessors
for each node as a heuristic when making chains, by choosing to add
the vertices with the fewest number of predecessors to a chain.
The algorithm has quadratic time complexity on the number
of tasks.

\section{ Priced Timed Automata and Priced Timed Markov Decision
  Processes}
\label{sec:Timed_Automata} Finding the optimal schedule in a task
graph can be reduced to location reachability in the theory of Timed
Automata~\cite{AlurDill1994,Abdeddaim2006SchedulingWithTA}. The
optimal schedule corresponds to the path with minimal time in the
corresponding automaton.
We extend the work in~\cite{Abdeddaim2006SchedulingWithTA} by using
the extensions of timed automata, we use Priced Timed Automata (PTA)
and Priced Timed Markov Decision Processes (PTMDP). Both PTA and
PTMPD, are useful for expressing task graph scheduling as the duration
of tasks can be expressed using clocks. Additionally by using prices,
we can optimize for the lowest total price, where the price could
represent the difficulty of execution. Finding the cheapest path,
based on the price variable, to a given goal location is also known as
the \textit{optimal reachability problem}, and when we add notions of
price and time it is denoted \textit{cost and time bounded optimal
  reachability problem}

In order to define how to model and solve task graph scheduling, we
define the minimal required syntax and semantics of PTA and PTMDP.
Our definitions are in the style of
\cite{Stratego_on_time_with_minimal_cost,Aceto:2007:RSM:1324845}.  We
refer the reader
to~\cite{Stratego_on_time_with_minimal_cost,Aceto:2007:RSM:1324845,Puterman:1994:MDP:528623}
for detailed definitions.

\subsection{Priced Timed Automata} In this extended version of Timed
Automata a price can be specified for staying in specific
locations. Prices are accumulated in a single continuous variable and
specifies a price per time unit for any given location of the
automaton.

\begin{definition}[PTA]
  \label{def:PricedTimedAuto} A PTA
  $\mathcal{A} = (L, l_0, X, \Sigma, E, P,Inv)$ is a tuple where $L$
  is a finite set of locations, $l_0 \in L$ is the initial location,
  $X$ is a finite set of non-negative real-valued clocks, $\Sigma$ is
  a finite set of actions,
  $E \subseteq L\times \mathcal{B}(X)\times \Sigma \times 2^X \times
  L$ is a finite set of edges, $P\ :\ L \rightarrow N$ assigns a
  price-rate to each location, and
  $Inv\ :\ L \rightarrow \mathcal{B}(X)$ sets an invariant for each
  location.
\end{definition}

Semantically, a PTA $\mathcal{A}$ is a priced transition system,
consisting of states, an initial state, a finite set of actions, and a
transition relation. The states comprise pairs $(l,v)$ with $l$ being
some location and $v$ being a corresponding clock valuation, such that
it fulfills the invariant in that location. The finite set of actions
correspond to $\Sigma$. Lastly, the transition relation is (similarly
to Timed Automata) defined as action transitions and delay
transitions, where in action transitions using an action we reach a
new state where the location is new but the clock has not
advanced. Naturally, the guard of the edge and the invariant of the
new location must be satisfied. For delay transitions we reach a new
state where the location remains the same but time delays. Again, the
invariant of the location must remain fulfilled. Hence, the price of
an action transition is 0, whereas the price of a delay transition is
proportional to the delay according to the price rate of the given
location.

A run through a PTA is an alternating sequence of action and delay
transitions. The length of a run can be described as the number of
action transitions, which is the \emph{logical} length, or by the
total time that has advanced through the delay transitions, which is
denoted \emph{metric} length.

\subsection{Priced Timed Markov Decision Processes}
PTMDP extends PTA with Markov Decision Processes. To define PTMDP
formally we must first understand the notion of a Priced Timed Game,
PTG; A PTG $\mathcal{G}$ is a PTA whose actions $\Sigma$ are divided
into controllable ($\Sigma_c$) and uncontrollable ($\Sigma_u$)
actions. We now define PTMDP formally, as we assume the choices of
delay and uncontrollable actions are stochastic and given according to
a (delay,action)-probability density function for a given state.

\begin{definition}[Priced Timed Markov Decision Processes]
  \label{def:PTMDP} A PTMDP is a pair $\mathcal{M} = (\mathcal{G},
\mu^u)$, where $\mathcal{G} = (L, l_0, X, \Sigma_c, \Sigma_u, E, P,
Inv)$ is a PTG, and $\mu^u$ is a family of density-functions,
$\{\mu^u_q : \exists l \exists v.q = (d, v)\}$, with $\mu^u_q(d, u)
\in\mathbb{R}_{\geq0}$ assigning the density of the environment aiming
at taking the uncontrollable action $u \in \Sigma_u$ after a delay of
$d$ from state $q$.
\end{definition}

Using the notion of PTMDPs the cost and timed bounded reachability
problem consists of finding a \emph{strategy} that will reach the
given goal location(s) within a given amount of time or
cost. Informally, for PTMDPs a strategy is a family of probability
density functions that assigns the density of the controller aiming at
taking the controllable action after a given delay from a given
state. Intuitively, the strategy is defined similarly to $\mu^u$ for
Definition \ref{def:PTMDP}, however assigning the density of the
controllable actions. For the formal definition of a strategy we refer
the reader to~\cite{Stratego_on_time_with_minimal_cost}. When the
environment is stochastic such a strategy can be determined using
machine learning, in particular reinforcement learning.


\section{Near Optimal Scheduling}
\label{sec:UppaalModels}

In this section we explain the methodology for modeling task graphs
with and without preemption. We model these using both PTA and
PTMDPs. We use \uppaalcora to implement PTA models and \stratego to
implement PTMDP models. \uppaal provides a C-like language, which
allows for the use of variables, loops, etc.\ in the
models. \uppaalcora uses PTA and solves the cost optimal reachability
problem using a branch and bound algorithm. Several strategies for
branching are available, for example \textit{smallest heuristic first}
and \textit{random optimal depth first} to obtain near-optimal
solutions and \textit{best first} for an optimal solution. \stratego
uses various types of reinforcement learning to create a strategy that
minimizes a time-bounded reachability
function~\cite{Learn_Smarter_Not_Harder}.

We model task graphs as chains rather than individual tasks. This
reduces the number of comparisons needed to know which tasks have
their dependencies fulfilled. Thus, by modeling chains, we can reduce
the computation time of creating schedules. The complexity of
pre-computing the chains is quadratic on the number of tasks. For our
models, chains can be computed fast and the time of computation is
negligible compared to the time of computing a scheduler.  Finally, we
allow preemption to occur a once one of the scheduled tasks has
finished executing.

\subsection{Non-Preemptive Models}

\begin{figure*} \centering
  \includegraphics[width=0.75\textwidth]{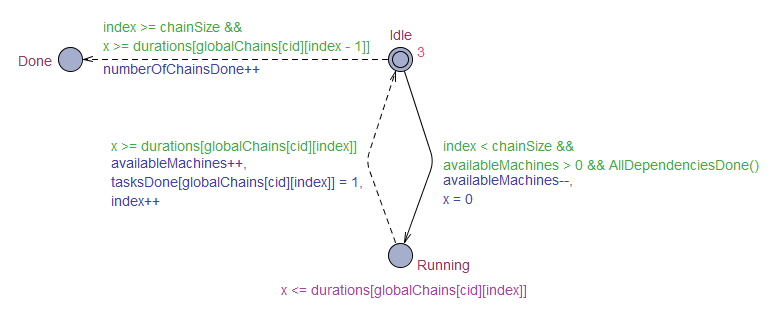}
  \caption{\stratego non-preemptive Chain template}
  \label{fig:StrategoNonPreemptive}
\end{figure*}

For the non-preemptive case, our \uppaalcora and our \stratego models
are alike and comprise two templates: \texttt{Composer} and
\texttt{Chain}. The \texttt{Composer} template is simple and is
therefore only described in text. The \texttt{Composer} has two
locations: \texttt{Init} and \texttt{Done}. The transition to
\texttt{Done} is enabled (and taken) when all tasks have finished
executing. The \texttt{Chain} template is seen in Figure
\ref{fig:StrategoNonPreemptive} and is created in \stratego. The
intuition of the \texttt{Chain} template is that it starts in
\texttt{Idle} and then takes the transition to \texttt{Running} to
execute a task. This is possible when a machine is available and a
task that is the least element of a chain has all its dependencies
completed. When the task has finished executing, the transition back
to idle is enabled and taken. This loop between \texttt{Idle} and
\texttt{Running} continues until all tasks in that chain finish
executing.

In \stratego transitions are either controllable or
uncontrollable. The controllable transitions denoted by solid arrows,
are the choices on which \stratego will use its machine learning
algorithms for optimization. Thus, in the \stratego model all
transitions are uncontrollable except the one from \texttt{Idle} to
\texttt{Running}. In this way, \stratego can decide and optimize when
to execute which tasks. We give the \texttt{Idle} state and the
\texttt{Init} state of the composer an exponential rate to increase
the probability of leaving the locations, thus ensuring advancement in
the model.

With \uppaalcora the \texttt{Chain} template is identical to
Figure~\ref{fig:StrategoNonPreemptive}, except all edges are
controllable and it has no exponential rate. The exponential rate is
not available in \uppaalcora as using the \texttt{cost’} variable
ensures advancement in the model. \uppaalcora models priced TA, so we
use the \texttt{cost’} variable in the composer template. For every
time unit spent in the \texttt{Init} location, the \texttt{cost’}
increases by one. This value is the one \uppaalcora minimizes in its
branch and bound algorithm.

\subsection{Preemptive Models}

\begin{figure*}[htb] \centering
  \includegraphics[width=0.75\textwidth]{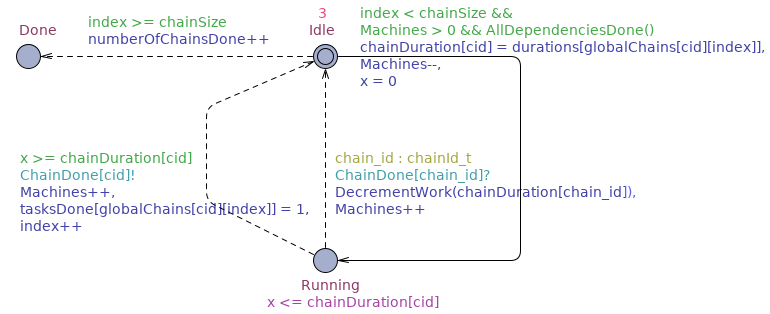}
  \caption{\stratego Preemptive Chain model}
  \label{fig:StrategoPreemptiveChain}
\end{figure*}

Our preemptive models share the structure of the non-preemptive, and
the \texttt{Composer} templates are identical. The preemptive
\texttt{Chain} template made in \stratego is seen in Figure
\ref{fig:StrategoPreemptiveChain}. Like in the non-preemptive case, it
comprises three locations being \texttt{Done}, \texttt{Idle}, and
\texttt{Running} and the general intuition remains the same. We
express the preemption in the two transitions from \texttt{Running} to
\texttt{Idle}. The leftmost is taken when a task has finished
executing, and it signals this to all other chains. All other chains
in \texttt{Running} receive the signal and preempt the task they were
running by taking the other transition from \texttt{Running} to
\texttt{Idle}. This behavior is implemented using \texttt{select}
statements and \texttt{broadcast} channels in \uppaal. The duration of
the preempted task is decremented with the duration of the task that
just finished executing. Following this, the model chooses a new set
of tasks to execute. As with the non-preemptive templates, the
\uppaalcora chain template differs by edges being controllable, having
no exponential rate, and adding the \texttt{cost’} variable.
%

\subsection{Correct Schedules}

We argue that the schedules computed using our UPPAAL models fulfill
the conditions of Definition \ref{def:Schedules}. To fulfill Condition
1 from Definition~\ref{def:Schedules} we store if a task is computed
using an array of booleans. We uphold the condition in the
\texttt{Chain} model on the guard of the transition from \texttt{Idle}
to \texttt{Running}. \texttt{AllDependenciesDone} computes
and-operations on the booleans in the array of computed tasks that
correspond to the indices of the task dependencies, resulting in
\textit{true} if the task can be scheduled. To ensure Condition 2 from
Definition~\ref{def:Schedules} the \texttt{Machines} counter
represents the number of available machines, initially being the total
number of machines. \texttt{Machines} increments when a task is
computed or preempted and decrements when a task is scheduled. The
guard on the transition from \texttt{Idle} to \texttt{Running}
requires that \texttt{Machines} must be positive, such that there is a
machine available. Our models fulfill Condition 3 from
Definition~\ref{def:Schedules}, as each task is executed in exactly
one chain and a chain can only compute one task at a time; the first
task in the chain. Lastly, Condition 4 from
Definition~\ref{def:Schedules} ensures that all tasks are fully
computed. As our models only compute the first task in the chain, this
fulfills the condition for all except the last task in each chain. For
the last task, the condition is fulfilled by the guard from
\texttt{Idle} to \texttt{Done}, where \texttt{index} must be greater
than or equal to \texttt{chainSize}. As the index variable is only
incremented when a task is complete, all tasks of the chain have been
computed when it is equal to the chain size. Additionally, no task can
be scheduled for longer than its duration because of the invariant in
the \texttt{Running} state.

%
\section{Experimental Evaluation}
\label{sec:tests_results}

In this section we evaluate our approach with an open data set
containing a large number of task graphs. We evaluate our models with
and without preemption, with and without chains. We compare our
results with an state of the art tool.

\subsection{Task Graph Set}
We use a standard task graph set of Kasahara and Narita
\cite{kasahara_schedule}, which contain task graphs of sizes from 50
to 5000 tasks, and for each task graph size the set contains 180
different task graphs. For each task graph, we give the shortest
obtained schedule length for 2, 4, 8, and 16 machines. The shortest
schedules are obtained from Kasahara and
Narita~\cite{kasahara_schedule} and are without preemption. Thus, our
preemptive runs \textit{can} have a lower absolute optimum. We test
our models on all 180 different task graphs for 50, 100, and 300
tasks, and extend the tests for \uppaalcora with the sizes 500, 750,
and 1000. We test each individual task graph with 2, 4, 8, and 16
machines both preemptively and non-preemptively in both \uppaalcora
and \stratego. In \uppaalcora, we use the branch and bound algorithm
for each task graph and number of machines. In each instance, we
conduct 100 runs with different pseudo-random seeds. To obtain
near-optimal solutions, we use the random optimal depth-first search
as the choice of branching. In \stratego, a strategy is learned and
the task graph is then simulated 2000 times under that strategy. All
schedules and their respective execution times are available on
GitHub\footnote{\url{https://github.com/marmux/spreadsheets}}.
\subsection{Experimental Setup}
We compare our schedules with the algorithm used by Kasahara and
Narita~\cite{kasahara_schedule}, it is a parallel depth first/implicit
heuristic search algorithm~\cite{Kasahara_article}.
We run each of our experiments on a single core of an AMD Opteron 6376
Processor. Aalborg University’s MCCAAU
cluster~\footnote{
\url{https://sites.google.com/site/mccaau/}
},
We are running Ubuntu 14.04.2 LTS with \uppaalcora 32bit,
and \stratego 4.1.20-stratego-5. 

To determine the quality of our resulting schedules, we compare these
to the best schedules from Kasahara and
Narita~\cite{kasahara_schedule}. We present this in
Table~\ref{tab:CoraResults} and Table~\ref{tab:StrategoResults}. For
Table~\ref{tab:StrategoResults} the task size ranges from 50 to 300
tasks, as with more \stratego consumes over 64 GB RAM, which we deem
too much for the experiments to continue. When computing the \stratego
preemptive models with 300 tasks, 27.6\% of the runs exceeded our
memory limit. For the results in Table~\ref{tab:CoraResults} the task
size ranges from 50 to 1000. \uppaalcora can at most use 4 GB of
RAM (32bit implementation), and we found that task graphs with over 360 chains
experience memory overflow, especially with more than four
machines. With 500 tasks the preemptive \uppaalcora model runs out of
memory in 1.24\% of the runs. For models with 750 tasks the preemptive
\uppaalcora runs out of memory in 5.08\% of the runs and the
non-preemptive model runs out of memory in 0.840\% of the
runs. Lastly, the task graphs with 1000 tasks, the memory overflow
occurred in 11.5\% of our preemptive runs and 1.37\% for our
non-preemptive runs.

\begin{table*}[!htb]
\centering
\caption{Aggregated results of \uppaalcora. \textit{\textbf{Size}} is the
  number of tasks in the task graphs being
  compared. \textit{\textbf{Mach.}} is the number of machines used for
  scheduling. Following is the percentage deviation between our
  schedules and the shortest-known schedules (from
  \cite{kasahara_schedule}); minimum length (\texttt{Min}), first
  quartile (\texttt{Q1}), median (\texttt{Q2}), third quartile
  (\texttt{Q3}), and maximum length (\texttt{Max}) of all the runs on
  the given task size and number of machines. Note, negative numbers
  denote instances where our schedules are shorter than those of
  \cite{kasahara_schedule}. We remove outliers greater than
  $Q3+(Q3-Q1) \cdot 2$ or less than $Q1-(Q3-Q1) \cdot 2$ to ensure an
  even distribution of data entries between the min, max, and
  quartiles.}
\label{tab:CoraResults}
\begin{tabularx}{\textwidth}{|c|c|Y|Y|Y|Y|c|Y|Y|Y|Y|Y|}
  \hline && \multicolumn{5}{c|}{\textit{\textbf{Cora Preemptive}}} &
  \multicolumn{5}{c|}{\textit{\textbf{Cora Non-preemptive}}}\\ \hline
  \textit{\textbf{Size}}& \textit{\textbf{Mach.}} &
  \textit{\textbf{Min}} & \textit{\textbf{Q1}} & \textit{\textbf{Q2}}
  & \textit{\textbf{Q3}} & \textit{\textbf{Max}} &
  \textit{\textbf{Min}} & \textit{\textbf{Q1}} & \textit{\textbf{Q2}}
  & \textit{\textbf{Q3}} & \textit{\textbf{Max}} \\\hline
  \multirow{4}{*}{\textit{50}}
  & 2 & -0.823 & 0.741 & 2.098 & 4.599 & 12.308 & 0.0 & 0.784 & 2.193 & 4.598 & 12.222 \\
  & 4 & 0.0 & 2.74 & 7.227 & 13.483 & 34.951 & 0.0 & 2.74 & 7.558 & 13.208 & 33.981 \\
  & 8 & -4.545 & 0.0 & 3.077 & 11.429 & 34.286 & 0.0 & 0.0 & 2.222 & 11.429 & 34.0 \\
  & 16 & 0.0 & 0.0 & 0.0 & 0.0 & 0.0 & 0.0 & 0.0 & 0.0 & 0.0 & 0.0 \\
  \hline \multirow{4}{*}{\textit{100}}
  & 2 & 0.0 & 0.432 & 1.228 & 3.372 & 9.227 & 0.0 & 0.541 & 1.345 & 3.303 & 8.824 \\
  & 4 & 0.0 & 2.308 & 4.803 & 8.917 & 22.059 & 0.0 & 2.21 & 4.478 & 8.333 & 20.513 \\
  & 8 & 0.0 & 0.0 & 4.706 & 11.765 & 35.088 & 0.0 & 0.0 & 3.141 & 10.145 & 30.435 \\
  & 16 & 0.0 & 0.0 & 0.0 & 1.709 & 5.128 & 0.0 & 0.0 & 0.0 & 0.0 & 0.0
  \\ \hline \multirow{4}{*}{\textit{300}}
  & 2 & 0.0 & 0.166 & 0.414 & 1.034 & 2.766 & 0.0 & 0.193 & 0.473 & 1.079 & 2.851 \\
  & 4 & 0.0 & 1.928 & 3.5 & 6.601 & 15.944 & 0.0 & 1.311 & 2.597 & 5.413 & 13.615 \\
  & 8 & 0.0 & 2.586 & 8.28 & 12.694 & 32.642 & 0.0 & 1.389 & 5.163 & 9.048 & 24.227 \\
  & 16 & 0.0 & 0.0 & 1.592 & 6.829 & 20.488 & 0.0 & 0.0 & 0.0 & 0.866
  & 2.571 \\ \hline \multirow{4}{*}{\textit{500}}
  & 2 & 0.0 & 0.077 & 0.224 & 0.571 & 1.555 & 0.0 & 0.114 & 0.259 & 0.578 & 1.5 \\
  & 4 & 0.0 & 1.408 & 2.395 & 4.888 & 11.835 & 0.0 & 0.73 & 1.42 & 3.29 & 8.41 \\
  & 8 & 0.0 & 4.502 & 8.644 & 12.352 & 28.008 & 0.0 & 1.791 & 4.159 & 7.692 & 19.336 \\
  & 16 & 0.0 & 0.548 & 3.367 & 12.874 & 37.5 & 0.0 & 0.0 & 0.0 & 3.917
  & 11.747 \\ \hline \multirow{4}{*}{\textit{750}}
  & 2 & 0.0 & 0.065 & 0.154 & 0.434 & 1.172 & 0.0 & 0.075 & 0.188 & 0.441 & 1.17 \\
  & 4 & 0.05 & 1.174 & 2.034 & 4.277 & 10.475 & 0.0 & 0.486 & 0.952 & 2.748 & 7.271 \\
  & 8 & 0.0 & 4.042 & 7.747 & 11.773 & 27.199 & 0.0 & 1.184 & 2.609 & 5.714 & 14.774 \\
  & 16 & 0.0 & 0.748 & 3.674 & 17.401 & 50.685 & 0.0 & 0.0 & 0.559 &
  5.93 & 17.742 \\ \hline \multirow{4}{*}{\textit{1000}}
  & 2 & 0.0 & 0.05 & 0.138 & 0.428 & 1.183 & 0.0 & 0.058 & 0.149 & 0.393 & 1.059 \\
  & 4 & 0.05 & 1.162 & 1.956 & 4.983 & 12.605 & 0.0 & 0.374 & 0.765 & 3.152 & 8.705 \\
  & 8 & 0.0 & 3.716 & 7.552 & 11.111 & 25.852 & 0.0 & 0.888 & 1.917 & 4.343 & 11.228 \\
  & 16 & 0.0 & 0.534 & 2.402 & 16.854 & 49.43 & 0.0 & 0.0 & 1.182 &
  5.814 & 17.431 \\ \hline
\end{tabularx}
\end{table*}

\begin{table*}[htb]
\centering
\caption{Aggregated results of Stratego. For an explanation of the contents, see Table \ref{tab:CoraResults}}
\label{tab:StrategoResults}
\begin{tabularx}{\textwidth}{|c|c|Y|Y|Y|Y|c|Y|Y|Y|Y|Y|}
\hline
&& \multicolumn{5}{c|}{\textit{\textbf{Stratego Preemptive}}} & \multicolumn{5}{c|}{\textit{\textbf{Stratego Non-preemptive}}}\\ \hline
\textit{\textbf{Size}}& \textit{\textbf{Mach.}} & \textit{\textbf{Min}} & \textit{\textbf{Q1}} & \textit{\textbf{Q2}} & \textit{\textbf{Q3}} & \textit{\textbf{Max}} &
  \textit{\textbf{Min}} & \textit{\textbf{Q1}} & \textit{\textbf{Q2}} & \textit{\textbf{Q3}} & \textit{\textbf{Max}} \\\hline
\multirow{4}{*}{\textit{50}}
& 2 & 0.0 & 0.0 & 0.0 & 0.0 & 0.0 & 0.0 & 0.0 & 0.0 & 0.727 & 2.174 \\
& 4 & -2.609 & 0.0 & 1.02 & 4.412 & 13.235 & -2.797 & 0.0 & 2.098 & 4.505 & 13.514 \\
& 8 & -8.791 & 0.0 & 0.0 & 4.651 & 13.953 & -2.326 & 0.0 & 0.0 & 2.083 & 6.25 \\
& 16 & 0.0 & 0.0 & 0.0 & 0.0 & 0.0 & 0.0 & 0.0 & 0.0 & 0.0 & 0.0 \\\hline
\multirow{4}{*}{\textit{100}}
& 2 & -0.654 & 0.0 & 0.0 & 0.364 & 1.091 & -0.24 & 0.0 & 0.321 & 0.743 & 2.23\\
& 4 & -0.508 & 1.042 & 2.817 & 5.224 & 13.455 & 0.0 & 1.071 & 2.655 & 4.592 & 11.628 \\ 
& 8 & -2.797 & 0.0 & 2.273 & 8.081 & 24.242 & 0.0 & 0.0 & 0.4 & 5.983 & 17.949 \\
& 16 & 0.0 & 0.0 & 0.0 & 0.0 & 0.0 & 0.0 & 0.0 & 0.0 & 0.0 & 0.0 \\\hline 
\multirow{4}{*}{\textit{300}} 
& 2 & 0.0 & 0.12 & 0.321 & 0.691 & 1.833 & 0.0 & 0.083 & 0.234 & 0.476 & 1.26 \\
& 4 & -0.127 & 1.099 & 2.228 & 4.674 & 11.808 & -0.127 & 1.027 & 1.921 & 3.3 & 7.843 \\
& 8 & 0.0 & 0.14 & 2.367 & 7.368 & 21.649 & 0.0 & 0.287 & 4.091 & 7.212 & 20.915 \\
& 16 & 0.0 & 0.0 & 0.0 & 0.0 & 0.0 & 0.0 & 0.0 & 0.0 & 0.218 & 0.654 \\\hline
\end{tabularx}
\end{table*}

\subsection{Discussion}
\label{sec:discussion}
In this section, we explain and discuss the results obtained in Table
\ref{tab:CoraResults} and Table \ref{tab:StrategoResults}. Because of
the large number of data points for each of the task graph sizes, we
have grouped all the data points by these sizes. Thus, we assume that
task graphs of the same size are equally challenging to find schedules
for. This is not necessarily true, as the state space of our models
grows with the number of chains rather than tasks. However, to express
as much of the data as possible, we use the minimum, quartiles, and
the maximum. These give a good notion of the distribution of the
values. Furthermore, we remove the outliers, which in most cases
removes the most radical one or two percent. The only issue with
removing the outliers occurs with the models where the first and the
third quartiles are equal. Here, we remove a larger number
(10-15\%). However, as all the remaining data values (85-90\%) have
the same value, we do not think this issue is significant. The results
of Table \ref{tab:CoraResults} and Table \ref{tab:StrategoResults}
contain a noticeable number of schedules that are as short as the
shortest-known \cite{kasahara_schedule}. This is because the optimal
non-preemptive schedule is found in both cases. Additionally, for each
task size, a minimum schedule length is obtained that is better than
or equal to those obtained in Kasahara and Narita
\cite{kasahara_schedule}. Looking at the tables, \stratego generally
obtains better schedules than \uppaalcora. In
Table~\ref{tab:StrategoResults}, we see that the preemptive models
perform better than the non-preemptive ones when considering the
minimum schedules. Furthermore, the preemptive models achieve the
highest maximum values. Scheduling preemptively leads to a much larger
state space, as many additional choices are available at any point
when scheduling. However, the overhead of the increased state space
does evidently not compromise the power of preemption to a significant
extent, especially when considering the minimum or the small task
graphs. Looking at the median values, the preemptive models generally
perform slightly worse than the non-preemptive, especially in the
cases with many tasks.

\begin{figure}[t]
    \centering
    \includegraphics[width=0.48\textwidth]{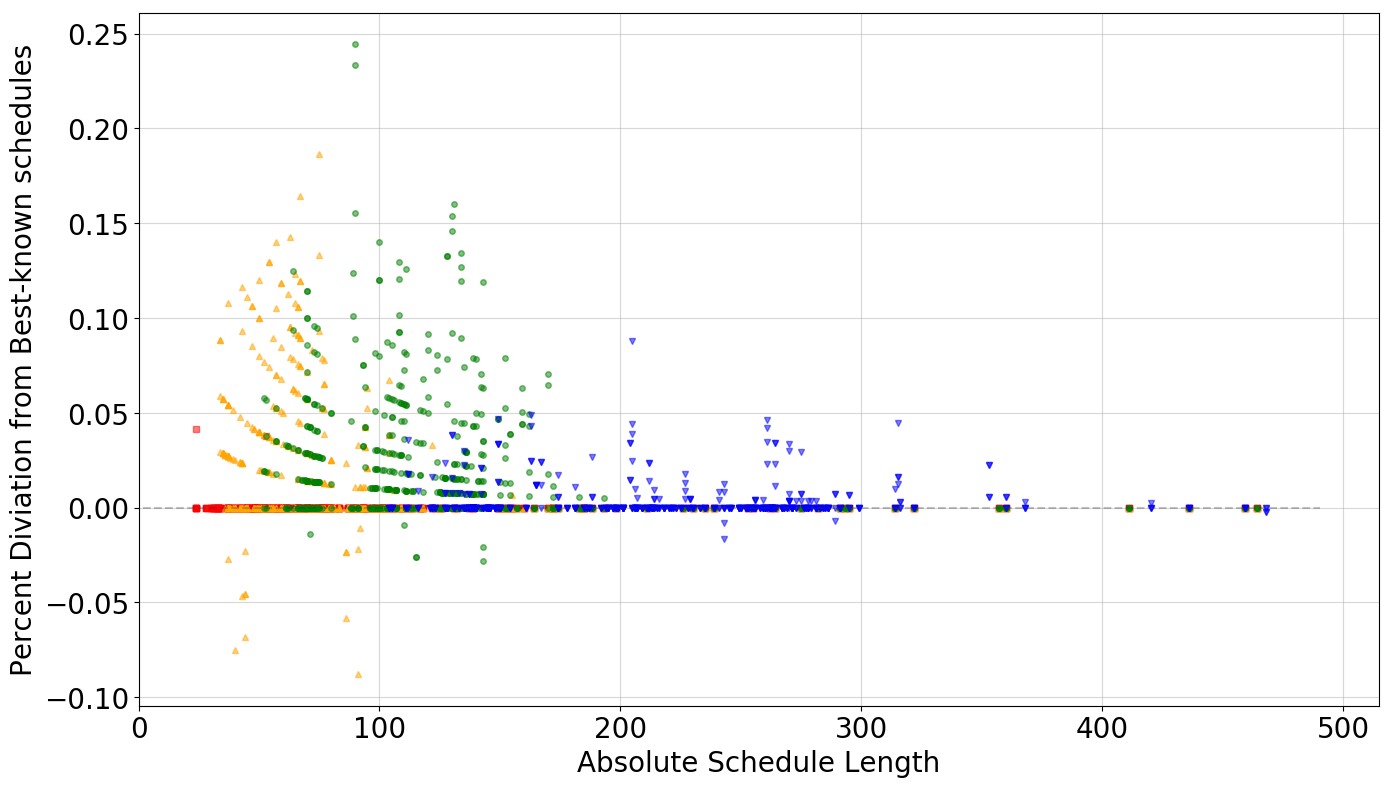}
    \caption{This plot aggregates the minimum schedules found for 50
      tasks for each model and number of machines. The colours and
      shapes separate results for 2, 4, 8, and 16 machines; blue
      triangles pointing down are 2, green circles are 4, orange
      triangles pointing up are 8, and red squares are 16. The x-axis
      is the length of the schedules we obtained, and the y-axis shows
      the relative deviation from the best-known schedules of Kasahara
      and Narita~\cite{kasahara_schedule}.}
    \label{fig:good_plot_50}
\end{figure}

Considering Figure~\ref{fig:good_plot_50} we see that the most
populated areas of the graph are on $y=0$, meaning that our schedules
match the known best schedules. Furthermore, $93\%$ of the entries where
the deviation is not $0\%$, is between $0\%$ and $10\%$ worse. Below
zero percent we see that some schedules found were better than those
found by Kasahara and Narita~\cite{kasahara_schedule}. Although there
are schedules up to $60\%$ worse, the area above 10\% is scarcely
populated considering the number of entries in the graphs. The large
deviations only occur on the shorter schedules, especially with a
large number of machines.

As we mention to in Section~\ref{sec:tests_results}, all the results
contain an execution time alongside the obtained schedule lengths. The
execution times show that \uppaalcora is significantly faster than
\stratego. Additionally, it is possible to run a single \uppaalcora
run, which provides a near-optimal schedule. This is not possible when
utilising machine learning in \stratego, as \stratego must make a
strategy before finding schedules. This is what most of the execution
time of \stratego is spent on. However, if the problem at hand is
static enough to reuse a strategy in \stratego, then the execution
time is significantly reduced. Additionally, \uppaalcora is limited to
4GB of RAM, whereas \stratego requires more RAM to compute the larger
task graphs. Thus, which model to use varies on the application.

\begin{figure}[t]
    \centering
    \includegraphics[width=0.48\textwidth]{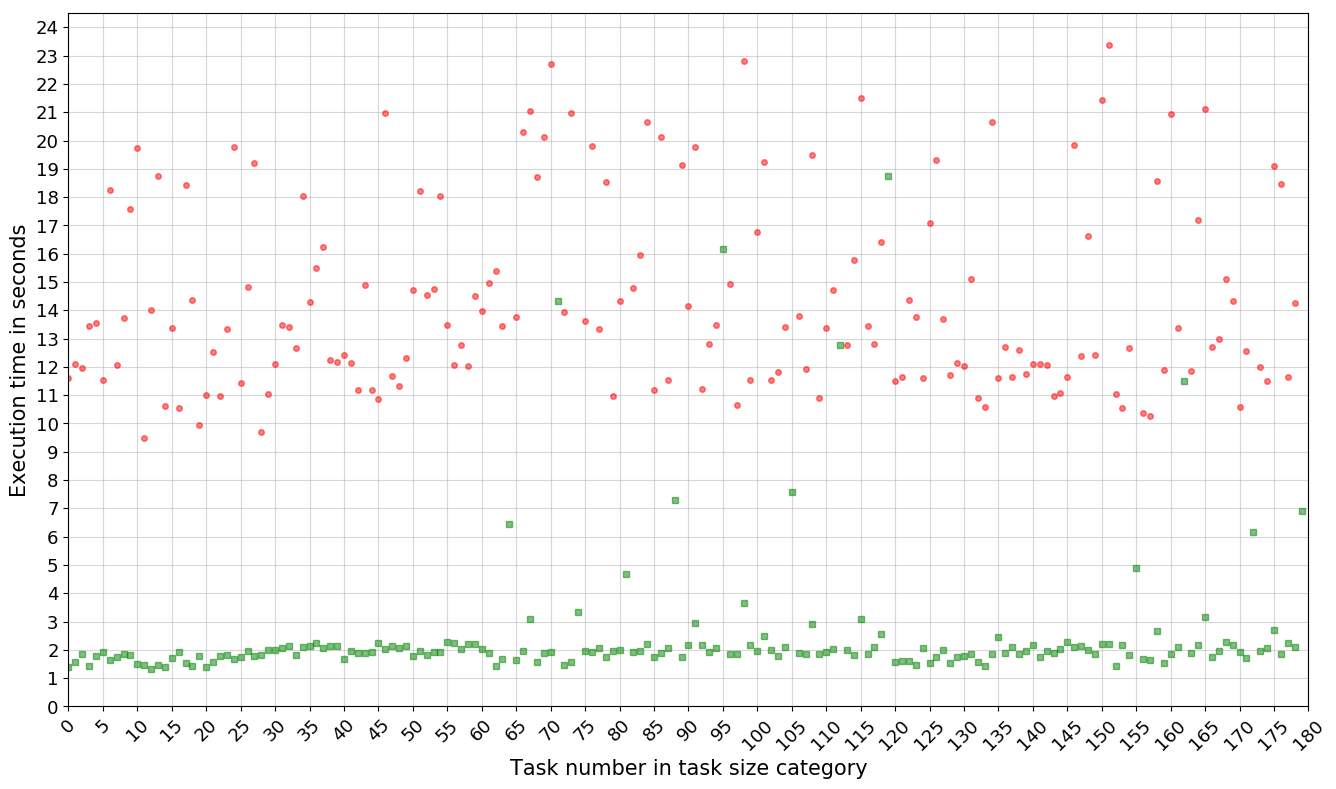}
    \caption{Execution time for non-preemptive \uppaalcora on the task-graphs
      of size 300, with 16 machines (Green squares represent runs with
      chains, red circles represent runs without chains)}
    \label{fig:executionTime}
\end{figure}

Furthermore, we timed the execution of all models with and without
chains. Figure~\ref{fig:executionTime} shows the execution time
between modelling the task graphs using chains and modelling all tasks
individually on the task size 300, with 16 machines for \uppaalcora
non-preemptive. On the figure, the red circles show the execution time
when modelling tasks individually, while green squares are runs where
chains are modelled. Note, that some of the results are not shown as
they are slower than 25 seconds. This is done to highlight the
difference in execution time. The only runs that were slower than 25
seconds were achieved when modelling tasks rather than
chains. Generally the execution time of \stratego is $\approx1.5$
times faster when using chains as compared to not using chains. For
\uppaalcora the is more substantial, as seen on Figure
\ref{fig:executionTime}, the execution time is $\approx8$ times faster
when using chains than without chains. This is highly generalised,
where smaller task graphs have smaller execution time differences,
than bigger task graphs.

In Figure~\ref{fig:good_plot_50} the minimum schedules found for 50
tasks for each model and number of machines are seen. We have made one
graph for each task size, and they are available on GitHub.  As only
\uppaalcora has run the task graphs of size 500 to 1000, these graphs
\textit{only} contain the 1440 data entries, while the other three
contain twice that. In contrast to Table \ref{tab:CoraResults} and
\ref{tab:StrategoResults} where we remove outliers, we do not remove
them in the plots. Thus, the results may seem to deviate. We choose
not to remove outliers in the plots, as we base these solely on the
minimum schedules obtained for each task graph and machine number.

\section{Conclusions and Future Work}
\label{sec:conclusion}

Task graph scheduling is a relevant problem in computer science with
application in diverse domains including production lines,
spreadsheets, etc. In this paper we present a methodology for finding
near-optimal schedulers for task graphs. We consider preemptive and
non-preemptive schedules. We model task graphs in the theories of
priced timed automata and priced timed markov decision processes. 
Our implementation uses \uppaalcora and \stratego to compute the
schedulers. We have compared our results with an state of the art
tool~\cite{kasahara_schedule}. Our experiments are encouraging and
show that in most models we perform better
than~\cite{kasahara_schedule}. We also explored the effect of using
chains, which proved to be beneficial.


Future work include syntactic optimizations on our models, for
example heuristics from~\cite{kasahara_schedule} could be adopted to
improve performance. Other future work include techniques to avoid the
state explosion problem explored. Partial Order Reduction (POR) has
been recently successfully applied to timed
systems~\cite{BJLMS:CAV:18}. Application of POR for timed systems with
costs could greatly improve the computation of schedulers for task
graphs.




\bibliography{articles}
\bibliographystyle{splncs04.bst}
\clearpage
\appendix

\section{Proofs}
\begin{proof}[Lemma \ref{lemma:Chain_optimality}]
  To prove Lemma \ref{lemma:Chain_optimality}, we show that a task is
  a least element of a chain at any point in time where it is
  scheduled. This proves Lemma \ref{lemma:Chain_optimality} as all
  tasks that can be scheduled at any point in time - meaning that
  their dependencies are fulfilled - will always be a least element of
  a chain at that point in time. Thus, at any point in time the set of
  all tasks that can be scheduled is a subset of the set of least
  elements in all chains. We show this using the conditions of
  Definition \ref{def:Schedules}.
  
  \label{proof:chains} Assume we have a task graph ($\mathcal{P}$,
  $\sqsubset$, D), a possible schedule $start$, and a chain
  decomposition: $C_1,C_2,...,C_k$ where $1\leq k\leq|\mathcal{P}|$. We
  denote the set of least elements in the chain decomposition at time
  $t$ as
  $ F = \{f\ |\ f\in C_i \setminus A\ for\ 1\leq i\leq
    k,\ where\ \forall x\in C_i \setminus A\ we\ have\ f\sqsubset x\ or\
    f=x\}$.
  Here $A$ is defined as the completed tasks at time
  $t$;
  \begin{equation} A = \{P\ \in \mathcal{P} |\ \sum_{\{(s',d') |
      (s',d') \in start(P)\ and\ s'+d' \leq t\}} d' = D(P)\}\ .
  \end{equation}

  Thus, to prove Lemma \ref{lemma:Chain_optimality} we need to show
  that; for any $P \in \mathcal{P}$ and for any $(s,d)\in start(P)$ then
  $P\in F$ at time s. We define the dependencies of $P$ as
  \begin{equation} Dep(P) = \{P' \in \mathcal{P}|P' \sqsubset P\}\ .
  \end{equation}

  From Condition 1 of Definition \ref{def:Schedules} we know that
  \begin{equation}
    \label{eq:DependencyFulfilled}
    \begin{array}{c} \forall P'\in Dep(P)\ s \geq sup(\{s'+d'\ |\
      (s',d')\in start(P')\})\ .
    \end{array}
  \end{equation} Recall, Condition 4 of Definition
  \ref{def:Schedules}, which states that the sum of all intermediate
  durations must be the sum of the full duration in a possible
  schedule. As the supremum of $s'+d' \in start(P')$ is less than or
  equal to $s$, as stated by equation \ref{eq:DependencyFulfilled}, then
  \begin{equation} \forall P'\in Dep(P)\ \sum_{\{(s',d') | (s',d') \in
      start(P')\ and\ s'+d' \leq s\}} d' = D(P')\ .
  \end{equation} Thus, $Dep(P) \subseteq A$. As every dependency of
  $P$ is in $A$, then $P$ cannot be dependent on any task in
  $C_i$. Furthermore, $P \in A$ can only happen in the case where $P$
  has finished computing and $d = 0$. This is not possible as
  $d\in\mathbb{R}_{>0}$, according to Definition
  \ref{def:Schedules}. Thus, $P \in F$.
\end{proof}

\section{Algorithm for Chain Decomposition}

The algorithm consists of two functions. The function
\texttt{ChainCover} first sorts the task graph into a topologically
sorted list. Then the function creates chains by calling
\texttt{Visit} until the sorted list is empty. \texttt{Visit} removes
the input node from the sorted list and calls itself recursively on
the successor with the smallest number of predecessors to follow the
heuristic. When a node is added to a chain all of its successors will
have their predecessor count decremented. This is done until the input
node has no successors and the entire chain is returned to the
\texttt{ChainCover} function. Once the sorted list is empty, then
\texttt{ChainCover} returns the list of chains. As \texttt{Visit} is
called exactly $|V|$ times and each call conciders at most $|V|$
successors, then the complexity of \texttt{ChainCover} is $O(|V|^2)$.

\begin{algorithm}
  \caption{Custom algorithm for computing the chain decomposition of a
task graph}
  \begin{algorithmic}[1] \Require A graph $G$, given by a list of
edges $E$, and a list of vertices $V$ \Ensure A chain decomposition of
the graph, given as a list of lists of vertices
\Function{ChainCover}{$G(E,V)$} \State SortedList $\gets$
Topologically\ sorted\ $G$ \State ChainsList $\gets$ NIL \ForAll{Edges
in $E$} \State Increment predecessorValue of destination vertex \State
Add destination vertex to the successors list of the source vertex
\EndFor \While{SortedList is not empty} \State Add chain from Visit
function to ChainsList, using the first node from SortedList and the
list itself as input.  \EndWhile \\ \Return ChainsList \EndFunction \\
\Function{Visit}{node,SortedList} \State Remove node from SortedList
\State Decrement predecessorValue of all successors of node \State
nextNode $\gets$ The successor with the smallest predecessorValue
\If{nextNode is NIL} \State \Return node \Else \State \Return
(node,visit(nextNode,SortedList) \EndIf \EndFunction
  \end{algorithmic}
\end{algorithm}



\end{document}